\documentclass[aps,pra,twocolumn,showpacs]{revtex4}%
\usepackage{graphics}
\usepackage{amsmath}
\usepackage{graphicx}
\usepackage{amsfonts}
\usepackage{amssymb}

\begin{document}
\title{Covariance matrices under Bell-like detections}
\author{Gaetana Spedalieri}
\author{Carlo Ottaviani}
\author{Stefano Pirandola}
\affiliation{Department of Computer Science, University of York, York YO10 5GH, United Kingdom}

\begin{abstract}
We derive a simple formula for the transformation of an arbitrary covariance
matrix of $(n+2)$ bosonic modes under general Bell-like detections, where the
last two modes are combined in an arbitrary beam splitter (i.e., with
arbitrary transmissivity) and then homodyned. In particular, we consider the
realistic condition of non-unit quantum efficiency for the homodyne detectors.
This formula can easily be specialized to describe the standard Bell
measurement and the heterodyne detection, which are exploited in many
contexts, including protocols of quantum teleportation, entanglement swapping
and quantum cryptography. In its general form, our formula can be adopted to
study quantum information protocols in the presence of experimental
imperfections and asymmetric setups, e.g., deriving from the use of unbalanced
beam splitters.

\end{abstract}

\pacs{03.67.--a, 02.10.Ud, 03.65.Ta}
\maketitle

\section{Introduction}

Gaussian quantum information is that area of quantum information which deals
with continuous variable systems (e.g., bosonic systems) prepared in Gaussian
states, evolving by Gaussian unitaries or channels, and finally measured by
Gaussian detectors~\cite{RMP}. Gaussian states are easy to generate
experimentally and very easy to manipulate theoretically. Their description
can be reduced to their first two statistical moments, which are the mean
value (or displacement vector) and the covariance matrix (CM). In particular,
the CM contains the most relevant information about the Gaussian state,
providing its entropy, purity properties and separability
properties~\cite{RMP}.

One of the most important Gaussian measurements is the Bell
detection~\cite{RMP,BraREV,BraREV2} (also known as continuous variable Bell
detection). This consists of combining two bosonic modes into a balanced-beam
splitter (i.e., with transmissivity $1/2$). The output modes are then measured
by two homodyne detectors in such a way that one mode is detected in the
position quadrature $\hat{q}$ and the other mode in the momentum quadrature
$\hat{p}$. This measurement is typical of a series of protocols with
continuous variable systems, including quantum
teleportation~\cite{Tele1,Tele2,Tele3,Tele4,Tele5,Tele6,TeleNET} and
entanglement swapping~\cite{entswap1,entswap2,entswap3,PeterENT}. Another
important measurement is heterodyne detection, where a single bosonic mode is
taken as input of a balanced-beam splitter (with the other input being the
vacuum) and the two outputs are homodyned in $\hat{q}$ and $\hat{p}$,
respectively. This is also a fundamental detection in many continuous variable
protocols, for instance in coherent-state quantum key
distribution~\cite{QKD1,QKD2,QKD3,QKD4} and two-way quantum
cryptography~\cite{2QKD}.

In this paper, we consider a generalized form of Bell measurement that we call
\textquotedblleft Bell-like detection\textquotedblright. Here we have two
bosonic modes which are combined into a beam splitter of \textit{arbitrary}
transmissivity $T$ and then homodyned in the two quadratures (one mode in
$\hat{q}$ and the other in $\hat{p}$). Standard Bell detection and heterodyne
detection are specific instances of this more general measurement. In our
derivation, we consider the general scenario where a set of $n+2$ bosonic
modes is given in a Gaussian state with arbitrary CM. By applying the
Bell-like detection to the last two modes of the set, we compute the
conditional reduced CM of the first $n$ modes surviving the measurement. This
is expressed in terms of the input CM\ and the beam splitter's transmissivity
$T$ adopted in the measurement. We derive this input-output formula both in
the ideal case of perfect detection, i.e., unit quantum efficiency for the
homodyne detectors, and the realistic case where detection is not necessarily
perfect, i.e., the homodyne detectors have arbitrary quantum efficiency
$0<\eta\leq1$ (a scenario which can be modelled by inserting additional beam
splitters in front of the detectors~\cite{Leonhardt}).

Our algebraic derivation is relatively easy starting from the well-known
transformation rules for CMs under partial homodyne
detections~\cite{Eisert,Eisert2}, which are here suitably generalized to the
case of arbitrary quantum efficiency $\eta$. Despite its easy derivation, our
main formula for Bell-like detections can be usefully applied in several
contexts. For instance, it can be exploited to extend the protocols of quantum
teleportation and entanglement swapping to considering unbalanced beam
splitters (asymmetric setups). Similarly, it can be used to perturb the ideal
model of heterodyne detection which is used in many protocols of quantum key distribution.

The paper is organized as follows. In Sec.~\ref{SEC1} we provide a brief
introduction to bosonic Gaussian states and CMs. In Sec.~\ref{SEC2} we review
the transformation rules for CMs under homodyne detections, generalizing these
well-known rules to the case of arbitrary quantum efficiency. Then, in
Sec.~\ref{SEC3}, we derive the main result of the paper, i.e., the formula for
the transformation of CMs under general Bell-like detections, which is first
given in the ideal case of unit efficiency and, then, in the general scenario
of arbitrary quantum efficiency for the homodyne detectors. Finally,
Sec.~\ref{SEC4} is for conclusions, with Appendix~\ref{Appendix} showing
specific examples of application of our results to the cases of standard Bell
detection and heterodyne detection.

\section{Basic notions on bosonic Gaussian states\label{SEC1}}

A system of $n$ bosonic modes is a described by a vector of $2n$ quadrature
operators%
\begin{equation}
\mathbf{\hat{x}}^{T}:=(\hat{q}_{1},\hat{p}_{1},\ldots,\hat{q}_{n},\hat{p}%
_{n})~,
\end{equation}
satisfying the commutation relations $[\hat{x}_{i},\hat{x}_{j}]=2i\Omega
_{ij}^{(n)}$, where $\Omega_{ij}^{(n)}$ is the generic element of the $n$-mode
symplectic form%
\begin{equation}
\mathbf{\Omega}^{(n)}=\bigoplus\limits_{k=1}^{n}\mathbf{\Omega~}%
,~\mathbf{\Omega:}=\left(
\begin{array}
[c]{cc}%
0 & 1\\
-1 & 0
\end{array}
\right)  ~. \label{Symplectic_Form}%
\end{equation}
By definition, a quantum state $\rho$ of $n$ bosonic modes is said to be
\textquotedblleft Gaussian\textquotedblright\ when its phase-space Wigner
function is Gaussian~\cite{RMP}. For this reason, a Gaussian state is fully
characterized by its first and second-order statistical moments. These are the
displacement vector%
\begin{equation}
\mathbf{\bar{x}}:=\mathrm{Tr}(\mathbf{\hat{x}}\rho)~,
\end{equation}
and the CM $\mathbf{V}$, with generic element%
\[
V_{ij}=\tfrac{1}{2}\mathrm{Tr}\left(  \{\hat{x}_{i},\hat{x}_{j}\}\rho\right)
-\bar{x}_{i},\bar{x}_{j}~.
\]
where $\{,\}$ denotes the anticommutator. By definition, the CM\ is a
$2n\times2n$ real and symmetric matrix. In order to be a quantum CM, it must
also satisfy the uncertainty principle~\cite{SIMONprinc}%
\begin{equation}
\mathbf{V}+i\mathbf{\Omega}^{(n)}\geq0~, \label{unc_PRINC}%
\end{equation}
or an equivalent set of bona-fide conditions (for instance, see
Ref.~\cite{TwomodePRA}\ for the case of two-mode CMs). In particular,
Eq.~(\ref{unc_PRINC}) implies the positivity definiteness
\begin{equation}
\mathbf{V}>0~.
\end{equation}
The simplest Gaussian state is the vacuum state, which corresponds to
$\mathbf{\bar{x}}=0$\ and $\mathbf{V}=\mathbf{I}$.

Once that a state is prepared in a Gaussian states, its evolution can be such
to preserve its Gaussian statistics. This is the case of Gaussian unitaries,
which are defined as those unitaries transforming Gaussian states into
Gaussian states. At the level of the second-order moments, the action of a
Gaussian unitary $\rho\rightarrow U\rho U^{\dagger}$\ corresponds to the
congruence transformation $\mathbf{V}\rightarrow\mathbf{SVS}^{T}$ where
$\mathbf{S}$ is a symplectic matrix, i.e., a matrix preserving the symplectic
form $\mathbf{S\Omega}^{(n)}\mathbf{S}^{T}=\mathbf{\Omega}^{(n)}$. A simple
example is the beam splitter transformation. This is defined by the
single-parameter symplectic matrix
\begin{equation}
\mathbf{K}(T)\mathbf{:=}\left(
\begin{array}
[c]{cc}%
\sqrt{T}\mathbf{I} & \sqrt{1-T}\mathbf{I}\\
-\sqrt{1-T}\mathbf{I} & \sqrt{T}\mathbf{I}%
\end{array}
\right)  ~, \label{BStrans}%
\end{equation}
where $\mathbf{I}$ is the $2\times2$ identity matrix and $0\leq T\leq1$ is the
transmissivity of the beam splitter. In the Heisenberg picture, the beam
splitter corresponds to the following Bogoliubov transformation of the
quadrature operators
\begin{equation}
\left(
\begin{array}
[c]{c}%
\hat{q}_{+}\\
\hat{p}_{+}\\
\hat{q}_{-}\\
\hat{p}_{-}%
\end{array}
\right)  =\mathbf{K}(T)\left(
\begin{array}
[c]{c}%
\hat{q}_{1}\\
\hat{p}_{1}\\
\hat{q}_{2}\\
\hat{p}_{2}%
\end{array}
\right)  =\left(
\begin{array}
[c]{c}%
\sqrt{T}\hat{q}_{1}+\sqrt{1-T}\hat{q}_{2}\\
\sqrt{T}\hat{p}_{1}+\sqrt{1-T}\hat{p}_{2}\\
-\sqrt{1-T}\hat{q}_{1}+\sqrt{T}\hat{q}_{2}\\
-\sqrt{1-T}\hat{p}_{1}+\sqrt{T}\hat{p}_{2}%
\end{array}
\right)  .
\end{equation}

Finally, Gaussian measurements can be defined as those quantum measurements
whose application to Gaussian states provides outcomes which are Gaussian
distributed~\cite{RMP}. When a Gaussian measurement is perfomed on a subset of
modes of a bosonic system prepared in a Gaussian state, then the reduced state
of the surviving (non-measured) modes is a Gaussian state. At the level of the
second-order moments, the CM of the final state is connected to the CM of the
initial state. As an example, when we homodyne one mode of a set of $n$
bosonic modes in a Gaussian state, the formula of the final CM has a
remarkably easy formula~\cite{Eisert,Eisert2}. This is reviewed in the next section.

\section{Covariance matrices under homodyne detections\label{SEC2}}

\subsection{Perfect homodyne detection}

Let us consider $n+1$ bosonic modes in a Gaussian state. This $(n+1)$-mode
Gaussian state $\rho_{in}$ has a CM that can be written in the blockform%
\begin{equation}
\mathbf{V}_{in}=\left(
\begin{array}
[c]{cc}%
\mathbf{A} & \mathbf{C}\\
\mathbf{C}^{T} & \mathbf{B}%
\end{array}
\right)  ~,
\end{equation}
where $\mathbf{A}$ is the CM of the first $n$ modes, $\mathbf{B}$ is the CM of
the last mode, and $\mathbf{C}$ is a rectangular $(2n\times2)$ real matrix
accounting for the cross correlations.

\begin{figure}[ptbh]
\vspace{-1.8cm}
\par
\begin{center}
\includegraphics[width=0.5\textwidth] {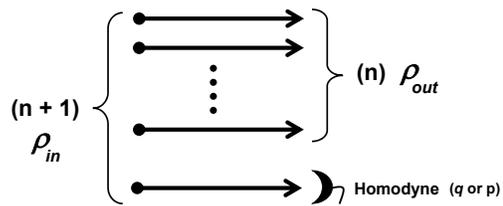}
\end{center}
\par
\vspace{-2.1cm}\caption{An input Gaussian state $\rho_{in}$ of $n+1$ modes is
homodyned in its last mode. The resulting output state $\rho_{out}$ of the
first $n$ modes is Gaussian. The input and output CMs are related by
Eq.~(\ref{Hom_law}) for $\hat{q}$-detection, and by Eq.~(\ref{Hom_law2}) for
$\hat{p}$-detection.}%
\label{CMpic}%
\end{figure}

Now, let us homodyne the $(n+1)^{\text{th}}$ mode as shown in Fig.~\ref{CMpic}%
, performing the detection of the $\hat{q}$ quadrature. The output state
$\rho_{out}$ of the remaining $n$ modes is still Gaussian. In particular, this
$n$-mode Gaussian state is described by the following CM~\cite{Eisert,Eisert2}%
\begin{equation}
\mathbf{V}_{out|q}=\mathbf{A-C}(\boldsymbol{\Pi}\mathbf{B}\boldsymbol{\Pi
})^{-1}\mathbf{C}^{T}~, \label{Hom_law}%
\end{equation}
where%
\begin{equation}
\boldsymbol{\Pi}:=\left(
\begin{array}
[c]{cc}%
1 & 0\\
0 & 0
\end{array}
\right)  ~.
\end{equation}
If we detect the $\hat{p}$\ quadrature, we have to consider the replacement%
\begin{equation}
\boldsymbol{\Pi}\rightarrow\boldsymbol{\Pi}^{\prime}:=\left(
\begin{array}
[c]{cc}%
0 & 0\\
0 & 1
\end{array}
\right)  ~, \label{Replac_P}%
\end{equation}
so that the conditional output CM\ is given by%
\begin{equation}
\mathbf{V}_{out|p}=\mathbf{A-C}(\boldsymbol{\Pi}^{\prime}\mathbf{B}%
\boldsymbol{\Pi}^{\prime})^{-1}\mathbf{C}^{T}~. \label{Hom_law2}%
\end{equation}
It is important to note that, in Eqs.~(\ref{Hom_law}) and~(\ref{Hom_law2}),
the matrices $\boldsymbol{\Pi}\mathbf{B}\boldsymbol{\Pi}$ and $\boldsymbol{\Pi
}^{\prime}\mathbf{B}\boldsymbol{\Pi}^{\prime}$ are singular, so that
$(\boldsymbol{\Pi}\mathbf{B}\boldsymbol{\Pi})^{-1}$ and $(\boldsymbol{\Pi
}^{\prime}\mathbf{B}\boldsymbol{\Pi}^{\prime})^{-1}$ must be interpreted as
pseudoinverses. In general, for a singular matrix $\mathbf{M}$, the
pseudoinverse $\mathbf{M}^{-1}$ (also known as Moore-Penrose inverse)\ is a
matrix $\mathbf{G}$ which minimizes the quantity
\[
r:=\sum\limits_{ij}(\mathbf{H}_{ij})^{2}\geq0~,
\]
where $\mathbf{H}_{ij}$ are the entries of $\mathbf{H:=MG-I}$, with
$\mathbf{I}$ being the identity matrix.

In the present problem, the pseudoinverses are easy to compute. In fact, let
us set
\begin{equation}
\mathbf{B}:=\left(
\begin{array}
[c]{cc}%
b_{1} & b_{3}\\
b_{3} & b_{2}%
\end{array}
\right)  ~, \label{Bform}%
\end{equation}
where $b_{1}>0$ and $b_{2}>0$, since $\mathbf{B}>0$ (being a reduced CM).
Then, we have $\boldsymbol{\Pi\mathbf{B\Pi}}=b_{1}\boldsymbol{\Pi}$, and we
can easily compute%
\begin{equation}
(\boldsymbol{\Pi}\mathbf{B}\boldsymbol{\Pi})^{-1}=\left(  b_{1}\boldsymbol{\Pi
}\right)  ^{-1}=(b_{1})^{-1}\boldsymbol{\Pi}~. \label{inv1}%
\end{equation}
This is a consequence of the fact we have $(x\boldsymbol{\Pi})^{-1}%
=x^{-1}\boldsymbol{\Pi}$ for any $x\neq0$~\cite{Proof}. Thus, for the $\hat
{q}$-detection, we can write%
\begin{equation}
\mathbf{V}_{out|q}=\mathbf{A-}(b_{1})^{-1}\mathbf{C}\boldsymbol{\Pi}%
\mathbf{C}^{T}~. \label{qsimple}%
\end{equation}
Similarly, for the detection of the other quadrature, we have $\boldsymbol{\Pi
}^{\prime}\boldsymbol{\mathbf{B\Pi}}^{\prime}=b_{2}\boldsymbol{\Pi}^{\prime}$,
so that%
\begin{equation}
(\boldsymbol{\Pi}^{\prime}\mathbf{B}\boldsymbol{\Pi}^{\prime})^{-1}%
=(b_{2})^{-1}\boldsymbol{\Pi}^{\prime}~. \label{inv2}%
\end{equation}
Thus, the formula for the $\hat{p}$-detection is simply given by%
\begin{equation}
\mathbf{V}_{out|p}=\mathbf{A-}(b_{2})^{-1}\mathbf{C}\boldsymbol{\Pi}^{\prime
}\mathbf{C}^{T}~. \label{psimple}%
\end{equation}

\subsection{Generalization to arbitrary quantum efficiency}

\begin{figure}[ptbh]
\vspace{-1.5cm}
\par
\begin{center}
\includegraphics[width=0.5\textwidth] {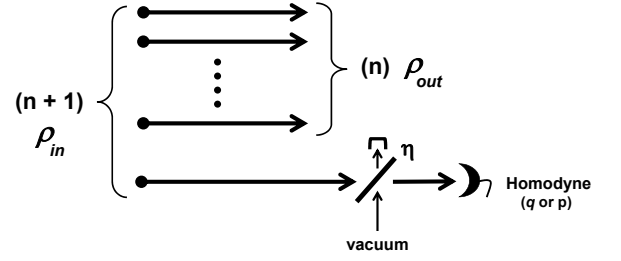}
\end{center}
\par
\vspace{-1.9cm}\caption{An input Gaussian state $\rho_{in}$ of $n+1$ modes is
homodyned in its last mode, with quantum efficiency $0<\eta\leq1$ (modelled as
a beam splitter of transmissivity $\eta$ which mixes the input signal mode
with an environmental vacuum mode). The resulting output state $\rho_{out}$ of
the first $n$ modes is Gaussian. The input and output CMs are related by
Eq.~(\ref{qsimpleEFF}) for $\hat{q}$-detection, and by Eq.~(\ref{psimpleEFF})
for $\hat{p}$-detection.}%
\label{small2}%
\end{figure}

Here we consider the case where the homodyne detector is not necessarily
perfect, i.e., it has a quantum efficiency $0<\eta\leq1$. This is modelled by
considering a beam-splitter with transmissivity $\eta$ in front of the
detector, where one port is accessed by the signal mode (the last mode of the
bosonic input set) and the other port is accessed by an environmental vacuum
mode. This scenario is depicted in Fig.~\ref{small2}

The generalization of the previous formulas is quite easy. The input CM is
first dilated to include the vacuum, i.e., $\mathbf{V}_{in}\rightarrow
\mathbf{V}^{\prime}:=\mathbf{V}_{in}\oplus\mathbf{I}$. Then, we apply the beam
splitter matrix to the last two modes, i.e.,
\begin{equation}
\mathbf{V}^{\prime}\rightarrow\mathbf{V}^{\prime\prime}:=[\mathbf{I}%
^{(n)}\oplus\mathbf{K}]\mathbf{V}^{\prime}[\mathbf{I}^{(n)}\oplus
\mathbf{K}]^{T}~,
\end{equation}
where $\mathbf{K}=\mathbf{K}(\eta)$ is given in Eq.~(\ref{BStrans}), and
\begin{equation}
\mathbf{I}^{(n)}=\bigoplus\limits_{k=1}^{n}\mathbf{I}%
\end{equation}
is the $n$-mode identity matrix ($2n\times2n$). The next step is to trace out
the transmission of the vacuum (i.e., the last output mode), which corresponds
to delete the last two rows and columns of the CM\ $\mathbf{V}^{\prime\prime}%
$. Thus, we have the following output CM for the $n+1$ bosonic modes before
detection%
\begin{equation}
\mathbf{V}^{\prime\prime\prime}=\left(
\begin{array}
[c]{cc}%
\mathbf{A} & \sqrt{\eta}\mathbf{C}\\
\sqrt{\eta}\mathbf{C}^{T} & \eta\mathbf{B}+(1-\eta)\mathbf{I}%
\end{array}
\right)  ~. \label{Vterza}%
\end{equation}
Note that the block $\mathbf{B}(\eta):=\eta\mathbf{B}+(1-\eta)\mathbf{I}$ is
positive-definite since it is the reduced CM of the last signal mode after the
beam-splitter. By expressing $\mathbf{B}$\ in the form of Eq.~(\ref{Bform}),
the diagonal terms of $\mathbf{B}(\eta)$ can be written as $b_{1}(\eta):=\eta
b_{1}+1-\eta>0$ and $b_{2}(\eta):=\eta b_{2}+1-\eta>0$.

Now, for $\hat{q}$-detection, we apply the formula of Eq.~(\ref{qsimple}) to
the CM of Eq.~(\ref{Vterza}). This is equivalent to make the replacements
$b_{1}\rightarrow b_{1}(\eta)$ and $\mathbf{C}\rightarrow\sqrt{\eta}%
\mathbf{C}$ in Eq.~(\ref{qsimple}). As a result, we get the final formula%
\begin{equation}
\mathbf{V}_{out|q}(\eta)=\mathbf{A-}\left(  b_{1}+\frac{1-\eta}{\eta}\right)
^{-1}\mathbf{C}\boldsymbol{\Pi}\mathbf{C}^{T}~, \label{qsimpleEFF}%
\end{equation}
for any quantum efficiency $0<\eta\leq1$. On the other hand, if we consider
the $\hat{p}$-detection, we apply the formula of Eq.~(\ref{psimple}) with
$b_{2}\rightarrow b_{2}(\eta)$ and $\mathbf{C}\rightarrow\sqrt{\eta}%
\mathbf{C}$. Thus, we find the other general formula%
\begin{equation}
\mathbf{V}_{out|p}(\eta)=\mathbf{A-}\left(  b_{2}+\frac{1-\eta}{\eta}\right)
^{-1}\mathbf{C}\boldsymbol{\Pi}^{\prime}\mathbf{C}^{T}~, \label{psimpleEFF}%
\end{equation}
for any quantum efficiency $0<\eta\leq1$.

\subsubsection{Example: Remote state preparation}

As a simple example of application, we consider the remote state preparation
which is typical in continuous variable quantum cryptography~\cite{RMP}. Alice
has an Einstein-Podolsky-Rosen (EPR) state~\cite{EPR}, which is a Gaussian
state with zero mean and CM equal to
\begin{equation}
\mathbf{V}_{\mathrm{EPR}}=\left(
\begin{array}
[c]{cc}%
\mu\mathbf{I} & \sqrt{\mu^{2}-1}\mathbf{Z}\\
\sqrt{\mu^{2}-1}\mathbf{Z} & \mu\mathbf{I}%
\end{array}
\right)  ~,
\end{equation}
with parameter $\mu\geq1$ and%
\begin{equation}
\mathbf{Z:=}\left(
\begin{array}
[c]{cc}%
1 & \\
& -1
\end{array}
\right)  ~. \label{ReflectionMAT}%
\end{equation}
Suppose that Alice measures the $\hat{q}$-quadrature of one mode with homodyne
efficiency $0<\eta\leq1$. Then, the other mode is projected into a Gaussian
state with CM
\begin{equation}
\mathbf{V}_{out|q}(\eta)=\mu\mathbf{I-}\frac{\mu^{2}-1}{\mu+\frac{1-\eta}%
{\eta}}\boldsymbol{\Pi}=\left(
\begin{array}
[c]{cc}%
\frac{\eta+(1-\eta)\mu}{\eta\mu+1-\eta} & \\
& \mu
\end{array}
\right)  ~.
\end{equation}
In particular, for $\eta=1/2$, we have
\begin{equation}
\mathbf{V}_{out|q}(\tfrac{1}{2})=\left(
\begin{array}
[c]{cc}%
1 & \\
& \mu
\end{array}
\right)  ~, \label{asyQ}%
\end{equation}
which is an asymmetric Gaussian state, with vacuum fluctuations in the
$\hat{q}$-quadrature and thermal in the $\hat{p}$-quadrature. In the case of
ideal detection $\eta=1$, we have%
\begin{equation}
\mathbf{V}_{out|q}(1)=\left(
\begin{array}
[c]{cc}%
\mu^{-1} & \\
& \mu
\end{array}
\right)  ~,
\end{equation}
which is the CM of a position-squeezed pure state.

Similarly, if Alice detects the $\hat{p}$-quadrature, we have%
\begin{equation}
\mathbf{V}_{out|p}(\eta)=\mu\mathbf{I-}\frac{\mu^{2}-1}{\mu+\frac{1-\eta}%
{\eta}}\boldsymbol{\Pi}^{\prime}=\left(
\begin{array}
[c]{cc}%
\mu & \\
& \frac{\eta+(1-\eta)\mu}{\eta\mu+1-\eta}%
\end{array}
\right)  ~.
\end{equation}
For $\eta=1/2$, Alice remotely prepares the other asymmetric Gaussian state%
\begin{equation}
\mathbf{V}_{out|p}(\tfrac{1}{2})=\left(
\begin{array}
[c]{cc}%
\mu & \\
& 1
\end{array}
\right)  ,
\end{equation}
while for ideal detection $\eta=1$, she remotely prepares a momentum-squeezed
pure state%
\begin{equation}
\mathbf{V}_{out|p}(1)=\left(
\begin{array}
[c]{cc}%
\mu & \\
& \mu^{-1}%
\end{array}
\right)  ~.
\end{equation}

\section{Covariance matrices under Bell-like detections\label{SEC3}}

In this section we derive the transformation rule for the CM\ under
generalized Bell-like detections, first assuming the condition of unit quantum
efficiency for the homodyne detectors (Sec.~\ref{IDEALbell}) and, then, the
general case of arbitrary quantum efficiencies~(Sec.~\ref{REALbell}).

\subsection{Ideal Bell-like measurements\label{IDEALbell}}

\begin{figure}[ptbh]
\vspace{-0.0cm}
\par
\begin{center}
\includegraphics[width=0.49\textwidth] {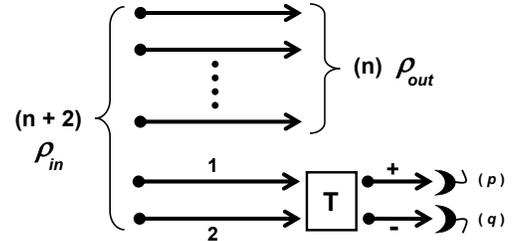}
\end{center}
\par
\vspace{-2.5cm}\caption{An input Gaussian state $\rho_{in}$ of $n+2$ modes is
subject to an ideal Bell-like detection (with arbitrary transmissivity $0\leq
T\leq1$) in the last two modes (labelled by $1$ and $2$). The output state
$\rho_{out}$ of the surviving $n$ modes is Gaussian. The output CM
$\mathbf{V}_{out}$ is related to the input CM $\mathbf{V}_{in}$ by
Eq.~(\ref{output_FORMULA}).}%
\label{CMpic2}%
\end{figure}

As depicted in Fig.~\ref{CMpic2}, let us consider $n+2$ bosonic modes in a
Gaussian state $\rho_{in}$. Its CM can always be written in the blockform%
\begin{equation}
\mathbf{V}_{in}=\left(
\begin{array}
[c]{cc}%
\mathbf{A} & \mathbf{C}\\
\mathbf{C}^{T} & \mathbf{B}^{(2)}%
\end{array}
\right)  ~, \label{Vin1}%
\end{equation}
where $\mathbf{A}$ is the reduced CM of the first $n$ modes,%
\begin{equation}
\mathbf{B}^{(2)}=\left(
\begin{array}
[c]{cc}%
\mathbf{B}_{1} & \mathbf{D}\\
\mathbf{D}^{T} & \mathbf{B}_{2}%
\end{array}
\right)  \label{Vin2}%
\end{equation}
is the reduced CM of the last two modes (labelled by $1$ and $2$), and%
\begin{equation}
\mathbf{C=}\left(
\begin{array}
[c]{cc}%
\mathbf{C}_{1} & \mathbf{C}_{2}%
\end{array}
\right)  =\left(
\begin{array}
[c]{cc}%
\vdots & \vdots\\
\mathbf{C}_{1}^{k} & \mathbf{C}_{2}^{k}\\
\vdots & \vdots
\end{array}
\right)  _{k=1,n} \label{Vin3}%
\end{equation}
is a rectangular $(2n\times4)$ real matrix, describing the correlations
between the first $n$ modes and the last two modes.

Here, we consider an ideal Bell-like detection applied to the last two modes
$1$ and $2$. This detection consists in applying a beam splitter of
transmissivity $T$, which transforms the input modes $1$ and $2$\ into the
output modes \textquotedblleft$+$\textquotedblright\ and \textquotedblleft%
$-$\textquotedblright, followed by two conjugate ($p$- and $q$-) homodyne
detections as shown in Fig.~\ref{CMpic2}. Thus, as a first step, let us apply
the beam-splitter symplectic matrix. The $(n+2)$-mode Gaussian state
$\tilde{\rho}$ at the output of the beam splitter has CM%
\begin{equation}
\mathbf{\tilde{V}}=[\mathbf{I}^{(n)}\oplus\mathbf{K}]\mathbf{V}_{in}%
[\mathbf{I}^{(n)}\oplus\mathbf{K}]^{T}~,
\end{equation}
where $\mathbf{I}^{(n)}$ is the $n$-mode identity matrix, and $\mathbf{K}%
=\mathbf{K}(T)$\ is the beam-splitter matrix of Eq.~(\ref{BStrans}) applied to
the last two modes. This CM takes the blockform%
\begin{equation}
\mathbf{\tilde{V}}=\left(
\begin{array}
[c]{cc}%
\mathbf{A} & \mathbf{\tilde{C}}\\
\mathbf{\tilde{C}}^{T} & \mathbf{\tilde{B}}^{(2)}%
\end{array}
\right)  ~, \label{Vtilde1}%
\end{equation}
where%
\begin{equation}
\mathbf{\tilde{C}}=\mathbf{CK}^{T}~,
\end{equation}
and%
\begin{equation}
\mathbf{\tilde{B}}^{(2)}=\mathbf{KB}^{(2)}\mathbf{K}^{T}~.
\end{equation}
More explicitly, the various blocks of the previous CM have the following
expressions%
\begin{equation}
\mathbf{\tilde{C}}=\left(
\begin{array}
[c]{cc}%
\mathbf{\tilde{C}}_{1} & \mathbf{\tilde{C}}_{2}%
\end{array}
\right)  =\left(
\begin{array}
[c]{cc}%
\vdots & \vdots\\
\mathbf{\tilde{C}}_{1}^{k} & \mathbf{\tilde{C}}_{2}^{k}\\
\vdots & \vdots
\end{array}
\right)  _{k=1,n}~,
\end{equation}
with%
\begin{align}
\mathbf{\tilde{C}}_{1}  &  =\sqrt{T}\mathbf{C}_{1}+\sqrt{1-T}\mathbf{C}%
_{2}~,\\
\mathbf{\tilde{C}}_{2}  &  =-\sqrt{1-T}\mathbf{C}_{1}+\sqrt{T}\mathbf{C}_{2}~,
\end{align}
and%
\begin{equation}
\mathbf{\tilde{B}}^{(2)}=\left(
\begin{array}
[c]{cc}%
\mathbf{\tilde{B}}_{1} & \mathbf{\tilde{D}}\\
\mathbf{\tilde{D}}^{T} & \mathbf{\tilde{B}}_{2}%
\end{array}
\right)  ~,
\end{equation}
with%
\begin{align}
\mathbf{\tilde{B}}_{1}  &  =T\mathbf{B}_{1}+(1-T)\mathbf{B}_{2}+\sqrt
{T(1-T)}(\mathbf{D}+\mathbf{D}^{T})~,\\
\mathbf{\tilde{B}}_{2}  &  =T\mathbf{B}_{2}+(1-T)\mathbf{B}_{1}-\sqrt
{T(1-T)}(\mathbf{D}+\mathbf{D}^{T})~,\\
\mathbf{\tilde{D}}  &  =\sqrt{T(1-T)}(\mathbf{B}_{2}-\mathbf{B}_{1}%
)+T\mathbf{D}-(1-T)\mathbf{D}^{T}~.
\end{align}
In terms of the previous blocks, the CM\ of Eq.~(\ref{Vtilde1}) takes the more
explicit form%
\begin{equation}
\mathbf{\tilde{V}}=\left(
\begin{array}
[c]{ccc}%
\mathbf{A} & \mathbf{\tilde{C}}_{1} & \mathbf{\tilde{C}}_{2}\\
\mathbf{\tilde{C}}_{1}^{T} & \mathbf{\tilde{B}}_{1} & \mathbf{\tilde{D}}\\
\mathbf{\tilde{C}}_{2}^{T} & \mathbf{\tilde{D}}^{T} & \mathbf{\tilde{B}}_{2}%
\end{array}
\right)  ~. \label{Vtilde2}%
\end{equation}
As already said, this CM\ describes the Gaussian state after the action of the
beam splitter which transforms the last two modes $1$ and $2$ into the output
modes \textquotedblleft$+$\textquotedblright\ and \textquotedblleft%
$-$\textquotedblright.

Now, we apply the $\hat{q}$-detection on the last mode \textquotedblleft%
$-$\textquotedblright, and the $\hat{p}$-detection on the next-to-last mode
\textquotedblleft$+$\textquotedblright. The detection of $\hat{q}_{-}$ implies
the transformation of Eq.~(\ref{Hom_law}), which here reads%
\begin{gather}
\mathbf{\tilde{V}}\rightarrow\mathbf{V}^{\prime}\nonumber\\
=\left(
\begin{array}
[c]{cc}%
\mathbf{A} & \mathbf{\tilde{C}}_{1}\\
\mathbf{\tilde{C}}_{1}^{T} & \mathbf{\tilde{B}}_{1}%
\end{array}
\right)  -\left(
\begin{array}
[c]{c}%
\mathbf{\tilde{C}}_{2}\\
\mathbf{\tilde{D}}%
\end{array}
\right)  \boldsymbol{\Gamma}\left(
\begin{array}
[c]{cc}%
\mathbf{\tilde{C}}_{2}^{T} & \mathbf{\tilde{D}}^{T}%
\end{array}
\right)  , \label{VcmTRA}%
\end{gather}
where%
\begin{equation}
\boldsymbol{\Gamma}:=(\boldsymbol{\Pi}\mathbf{\tilde{B}}_{2}\boldsymbol{\Pi
})^{-1}~. \label{gammadef}%
\end{equation}
In other words, after the detection of \textquotedblleft$-$\textquotedblright,
the $(n+1)$-mode CM describing the first $n$ modes and mode \textquotedblleft%
$+$\textquotedblright\ is given by%
\begin{equation}
\mathbf{V}^{\prime}=\left(
\begin{array}
[c]{cc}%
\mathbf{A}^{\prime} & \mathbf{C}^{\prime}\\
\mathbf{C}^{\prime T} & \mathbf{B}^{\prime}%
\end{array}
\right)  ~, \label{vprimoP}%
\end{equation}
where%
\begin{gather}
\mathbf{A}^{\prime}=\mathbf{A-\tilde{C}}_{2}~\boldsymbol{\Gamma~}%
\mathbf{\tilde{C}}_{2}^{T}~,\\
\mathbf{B}^{\prime}=\mathbf{\tilde{B}}_{1}-\mathbf{\tilde{D}~}%
\boldsymbol{\Gamma~}\mathbf{\tilde{D}}^{T}~,
\end{gather}
and%
\begin{equation}
\mathbf{C}^{\prime}=\mathbf{\tilde{C}}_{1}-\mathbf{\tilde{C}}_{2}%
~\boldsymbol{\Gamma~}\mathbf{\tilde{D}}^{T}~. \label{cprimoBLOCK}%
\end{equation}
Now, let us apply the $\hat{p}$-detection on mode \textquotedblleft%
$+$\textquotedblright. By using Eq.~(\ref{Hom_law2}), we get the final CM for
the first $n$ modes after the measurement, which is given by%
\begin{equation}
\mathbf{V}^{\prime}\rightarrow\mathbf{V}_{out}=\mathbf{A}^{\prime}%
-\mathbf{C}^{\prime}~\boldsymbol{\Gamma}^{\prime}~\mathbf{C}^{\prime T}~,
\label{outputCM}%
\end{equation}
where%
\begin{equation}
\boldsymbol{\Gamma}^{\prime}:=(\boldsymbol{\Pi}^{\prime}\mathbf{B}^{\prime
}\boldsymbol{\Pi}^{\prime})^{-1}~. \label{gammaPdef}%
\end{equation}

\subsubsection{Simplification of the input-output formula}

Here we simplify the formula for the output CM\ given in Eq.~(\ref{outputCM}).
Let us explicitly write the reduced CM\ $\mathbf{B}^{(2)}$ of the detected
modes by setting%
\begin{gather}
\mathbf{B}_{1}:=\left(
\begin{array}
[c]{cc}%
\beta_{1} & \beta_{3}\\
\beta_{3} & \beta_{2}%
\end{array}
\right)  ~,~\mathbf{B}_{2}:=\left(
\begin{array}
[c]{cc}%
\beta_{1}^{\prime} & \beta_{3}^{\prime}\\
\beta_{3}^{\prime} & \beta_{2}^{\prime}%
\end{array}
\right)  ~,\label{para1}\\
\mathbf{D}:=\left(
\begin{array}
[c]{cc}%
\delta_{1} & \delta_{3}\\
\delta_{4} & \delta_{2}%
\end{array}
\right)  ~.\label{para2}%
\end{gather}
From these matrices, we can construct the following real symmetric matrix%
\begin{equation}
\boldsymbol{\gamma}:=\left(
\begin{array}
[c]{cc}%
\gamma_{1} & \gamma_{3}\\
\gamma_{3} & \gamma_{2}%
\end{array}
\right)  ~,\label{gmatrix}%
\end{equation}
where%
\begin{align}
\gamma_{1} &  :=(1-T)\beta_{1}+T\beta_{1}^{\prime}-2\sqrt{T(1-T)}\delta
_{1}~,\label{gamma1}\\
\gamma_{2} &  :=T\beta_{2}+(1-T)\beta_{2}^{\prime}+2\sqrt{T(1-T)}\delta
_{2}~,\label{gamma2}%
\end{align}
and%
\begin{equation}
\gamma_{3}:=\sqrt{T(1-T)}(\beta_{3}^{\prime}-\beta_{3})-(1-T)\delta
_{3}+T\delta_{4}~.\label{gamma3}%
\end{equation}
Then, after simple algebra we get%
\begin{equation}
\boldsymbol{\Gamma}=\frac{\boldsymbol{\Pi}}{\gamma_{1}}\boldsymbol{~}%
,\label{gammaFOR}%
\end{equation}
and%
\begin{equation}
\boldsymbol{\Gamma}^{\prime}=\frac{\gamma_{1}}{\det\boldsymbol{\gamma}%
}\boldsymbol{\Pi}^{\prime}~.\label{gammaFOR2}%
\end{equation}
Note that previous Eqs.~(\ref{gammaFOR}) and~(\ref{gammaFOR2}) are
well-defined, since $\gamma_{1}>0$ and $\det\boldsymbol{\gamma}>0$, i.e., the
matrix $\boldsymbol{\gamma}$\ is positive definite. Using Eqs.~(\ref{gammaFOR}%
) and~(\ref{gammaFOR2}), we can simplify the previous Eq.~(\ref{outputCM}).
After some algebra, we get the first main result of our paper, i.e., the
input-output formula for the CM\ under ideal Bell-like detection%
\begin{equation}
\mathbf{V}_{out}=\mathbf{A}-\frac{1}{\det\boldsymbol{\gamma}}\sum_{i,j=1}%
^{2}\mathbf{C}_{i}\mathbf{K}_{ij}\mathbf{C}_{j}^{T}~,\label{output_FORMULA}%
\end{equation}
where%
\begin{align}
\mathbf{K}_{11} &  =\left(
\begin{array}
[c]{cc}%
(1-T)\gamma_{2} & \sqrt{T(1-T)}\gamma_{3}\\
\sqrt{T(1-T)}\gamma_{3} & T\gamma_{1}%
\end{array}
\right)  ,\label{KAPPA11}\\
\mathbf{K}_{22} &  =\left(
\begin{array}
[c]{cc}%
T\gamma_{2} & -\sqrt{T(1-T)}\gamma_{3}\\
-\sqrt{T(1-T)}\gamma_{3} & (1-T)\gamma_{1}%
\end{array}
\right)  ,\label{KAPPA22}\\
\mathbf{K}_{12} &  =\mathbf{K}_{21}^{T}=\left(
\begin{array}
[c]{cc}%
-\sqrt{T(1-T)}\gamma_{2} & (1-T)\gamma_{3}\\
-T\gamma_{3} & \sqrt{T(1-T)}\gamma_{1}%
\end{array}
\right)  .\label{KAPPA12}%
\end{align}
Thus, the output CM\ $\mathbf{V}_{out}$ of the surviving $n$ modes after the
ideal Bell-like detection is related to the input CM $\mathbf{V}_{in}$ of the
initial $n+2$ modes of Eqs.~(\ref{Vin1})-(\ref{Vin3}) by means of the
input-output relation of Eq.~(\ref{output_FORMULA}), where the matrices
$\mathbf{K}_{ij}$\ and $\boldsymbol{\gamma}$\ are completely characterized by
the reduced CM\ $\mathbf{B}^{(2)}$ of the detected modes and the transmission
$0\leq T\leq1$ which is used in the Bell-like detection.

Note that, in Eq.~(\ref{output_FORMULA}), the terms $\mathbf{C}_{i}%
\mathbf{K}_{ij}\mathbf{C}_{j}^{T}$ generate $2n\times2n$ matrices, i.e., with
the same dimensions of $\mathbf{A}$. For instance, we have%
\begin{gather}
\mathbf{C}_{1}\mathbf{K}_{11}\mathbf{C}_{1}^{T}=\left(
\begin{array}
[c]{c}%
\vdots\\
\mathbf{C}_{1}^{k}\\
\vdots
\end{array}
\right)  \mathbf{K}_{11}\left(
\begin{array}
[c]{ccc}%
\cdots & (\mathbf{C}_{1}^{k})^{T} & \cdots
\end{array}
\right) \nonumber\\
=\left(
\begin{array}
[c]{ccc}%
\mathbf{C}_{1}^{1}\mathbf{K}_{11}(\mathbf{C}_{1}^{1})^{T} & \cdots &
\mathbf{C}_{1}^{1}\mathbf{K}_{11}(\mathbf{C}_{1}^{n})^{T}\\
\vdots & \ddots & \vdots\\
\mathbf{C}_{1}^{n}\mathbf{K}_{11}(\mathbf{C}_{1}^{1})^{T} & \cdots &
\mathbf{C}_{1}^{n}\mathbf{K}_{11}(\mathbf{C}_{1}^{n})^{T}%
\end{array}
\right)  ~.
\end{gather}
As an exercise, we specify our ideal input-output formula of
Eq.~(\ref{output_FORMULA}) to the cases of standard Bell measurement and
heterodyne detection in Appendix~\ref{Appendix}.

\subsection{Bell-like measurements with arbitrary quantum
efficiencies\label{REALbell}}

\begin{figure}[ptbh] \vspace{-0.4cm}
\par
\begin{center}
\includegraphics[width=0.5\textwidth] {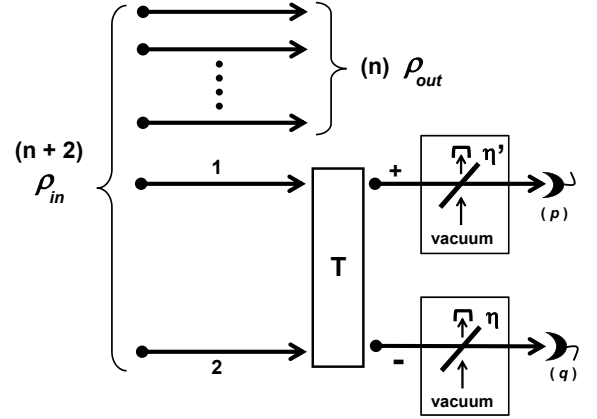}
\end{center}
\par
\vspace{-0.6cm}\caption{An input Gaussian state $\rho_{in}$ of $n+2$ modes is
subject to a realistic Bell-like detection in its last two modes (labelled by
$1$ and $2$). Homodyne detectors have arbitrary quantum efficiencies
$0<\eta\leq1$ and $0<\eta^{\prime}\leq1$ (modelled by beam-splitters with
transmissivities $\eta$ and $\eta^{\prime}$, which mix the last two signal
modes with two environmental vacua). The output state $\rho_{out}$ of the
surviving $n$ modes is Gaussian and its CM is related to the input CM by the
formula of Eq.~(\ref{newInOut}).}%
\label{largePIC}%
\end{figure}

In this subsection, we generalize the previous input-output
formula for the CM given in Eq.~(\ref{output_FORMULA}) to
realistic detectors. As depicted in Fig.~\ref{largePIC}, we
consider two homodyne detectors with quantum efficiencies
$0<\eta\leq1$ and $0<\eta^{\prime}\leq1$, modelled by inserting
two beam-splitters with transmissivities $\eta$ and
$\eta^{\prime}$, which mix the last two signal modes with
environmental vacua.

As before we start from the input CM $\mathbf{V}_{in}$\ of Eq.~(\ref{Vin1}),
whose blocks are specified in Eqs.~(\ref{Vin2}) and~(\ref{Vin3}), and their
parametrization is given in Eqs.~(\ref{para1}) and~(\ref{para2}). From this
parametrization, we construct the $\boldsymbol{\gamma}$-matrix of
Eq.~(\ref{gmatrix})\ as before, i.e., using Eqs.~(\ref{gamma1}),
(\ref{gamma2}) and~(\ref{gamma3}). In order to compute the new input-output
formula, note that the derivation is the same as before up to
Eq.~(\ref{Vtilde2}), which represents the CM\ of the state after the action of
the Bell's beam-splitter with transmissivity $T$. The difference is that we
now apply an imperfect $\hat{q}$-detection on the last mode \textquotedblleft%
$-$\textquotedblright\ with efficiency $\eta$, and an imperfect $\hat{p}%
$-detection on the next-to-last mode \textquotedblleft$+$\textquotedblright%
\ with efficiency $\eta^{\prime}$. This means that the steps from
Eq.~(\ref{VcmTRA}) to Eq.~(\ref{gammaPdef}) are still valid, proviso that we
suitably replace the two matrices $\boldsymbol{\Gamma}$ and
$\boldsymbol{\Gamma}^{\prime}$.

The inefficient $\hat{q}$-detection on the last mode \textquotedblleft%
$-$\textquotedblright\ corresponds to apply the transformation of
Eq.~(\ref{qsimpleEFF}) to Eq.~(\ref{Vtilde2}). As a result, we get
Eq.~(\ref{VcmTRA}) with%
\begin{equation}
\boldsymbol{\Gamma}=\frac{\boldsymbol{\Pi}}{\gamma_{1}(\eta)}~,\label{newGm}%
\end{equation}
where
\begin{equation}
\gamma_{1}(\eta):=\gamma_{1}+\frac{1-\eta}{\eta}~.\label{gamma1primo}%
\end{equation}
This expression clearly coincides with that of Eq.~(\ref{gammaFOR}) for unit
efficiency ($\eta=1$). Thus, the $(n+1)$-mode CM describing the first $n$
modes and mode \textquotedblleft$+$\textquotedblright\ is now given by
Eqs.~(\ref{vprimoP}-\ref{cprimoBLOCK}) with $\boldsymbol{\Gamma}$ defined in
Eq.~(\ref{newGm}).

We now apply the inefficient $\hat{p}$-detection on mode \textquotedblleft%
$+$\textquotedblright, which corresponds to apply the transformation of
Eq.~(\ref{psimpleEFF}) to Eq.~(\ref{vprimoP}). As a result we get the final CM
$\mathbf{V}_{out}$ for the first $n$ modes after the inefficient Bell-like
measurement, which is given by Eq.~(\ref{outputCM}) where the matrix
$\boldsymbol{\Gamma}^{\prime}$ is now equal to%
\begin{equation}
\boldsymbol{\Gamma}^{\prime}=\frac{\gamma_{1}(\eta)}{\det\boldsymbol{\gamma
}(\eta,\eta^{\prime})}\boldsymbol{\Pi}^{\prime}~,\label{newGm2}%
\end{equation}
where%
\begin{equation}
\boldsymbol{\gamma}(\eta,\eta^{\prime}):=\left(
\begin{array}
[c]{cc}%
\gamma_{1}(\eta) & \gamma_{3}\\
\gamma_{3} & \gamma_{2}(\eta^{\prime})
\end{array}
\right)  ~,\label{gammaPmatrix}%
\end{equation}
with $\gamma_{1}(\eta)$ defined in Eq.~(\ref{gamma1primo}) and%
\begin{equation}
\gamma_{2}(\eta^{\prime}):=\gamma_{2}+\frac{1-\eta^{\prime}}{\eta^{\prime}%
}~.\label{gamma2primo}%
\end{equation}
In other words, we can write
\begin{equation}
\boldsymbol{\gamma}(\eta,\eta^{\prime})=\boldsymbol{\gamma}+\boldsymbol{\Phi
}(\eta,\eta^{\prime})~,\label{gammaPmatrix2}%
\end{equation}
where
\begin{equation}
\boldsymbol{\Phi}(\eta,\eta^{\prime}):=\left(
\begin{array}
[c]{cc}%
\frac{1-\eta}{\eta} & \\
& \frac{1-\eta^{\prime}}{\eta^{\prime}}%
\end{array}
\right)  ~.\label{FiGrande}%
\end{equation}
The new matrix $\boldsymbol{\gamma}(\eta,\eta^{\prime})$ is essentially the
old matrix $\boldsymbol{\gamma}$ plus the effect $\boldsymbol{\Phi}(\eta
,\eta^{\prime})$ of the quantum efficiencies $\eta$ and $\eta^{\prime}$.
Again, for ideal detection ($\eta=\eta^{\prime}=1$) we have
$\boldsymbol{\gamma}(1,1)=\boldsymbol{\gamma}$ which means that
Eq.~(\ref{newGm2}) becomes identical to the previous Eq.~(\ref{gammaFOR2}).

Using Eqs.~(\ref{newGm}) and~(\ref{newGm2}) in Eq.~(\ref{outputCM}), we derive
the explicit expression of $\mathbf{V}_{out}$. After some algebra, we get the
second main result of our paper, i.e., the general input-output formula for
the CM under Bell-like detection with arbitrary quantum efficiencies $\eta$
and $\eta^{\prime}$. This is given by%
\begin{equation}
\mathbf{V}_{out}(\eta,\eta^{\prime})=\mathbf{A}-\frac{1}{\det
\boldsymbol{\gamma}(\eta,\eta^{\prime})}\sum_{i,j=1}^{2}\mathbf{C}%
_{i}\mathbf{K}_{ij}(\eta,\eta^{\prime})\mathbf{C}_{j}^{T}~,\label{newInOut}%
\end{equation}
where the matrices $\mathbf{K}_{ij}(\eta,\eta^{\prime})$ are equal to those
given in Eqs.~(\ref{KAPPA11}-\ref{KAPPA12}) up to the replacements
\begin{equation}
\gamma_{1}\rightarrow\gamma_{1}(\eta),~\gamma_{2}\rightarrow\gamma_{2}%
(\eta^{\prime}).\label{Replaces}%
\end{equation}

Note that the only difference between the general formula of
Eq.~(\ref{newInOut}) and the ideal formula of Eq.~(\ref{output_FORMULA}) is in
the replacements of Eq.~(\ref{Replaces}). In other words, to compute the
output CM, we perform exactly the same procedure as before for the ideal case,
proviso that we now use $\gamma_{1}(\eta)$ and $\gamma_{2}(\eta^{\prime})$ in
the $\boldsymbol{\gamma}$-matrix and the $\mathbf{K}$-matrices. As an
exercise, we specify our general input-output formula of Eq.~(\ref{newInOut})
to the cases of standard Bell measurement and heterodyne detection in
Appendix~\ref{Appendix}.

\section{Conclusion\label{SEC4}}

In conclusion, we have derived a simple formula for the transformation of CMs
under generalized Bell-like detections, where two modes of a bosonic system
are subject to an arbitrary beam-splitter transformation, followed by homodyne
detections. We have consider first the case of ideal detection and, then, the
general case of homodyne detectors with arbitrary quantum efficiencies. Our
formula can be applied to study quantum information protocols in various
contexts, including protocols of teleportation, entanglement swapping and
quantum key distribution. In particular, it can be adopted to generalize the
analysis of these protocols to the presence of experimental imperfections and
asymmetric setups, for instance, deriving from the use of unbalanced beam splitters.

\section{Acknowledgements}

This work has been supported by EPSRC under the research grant HIPERCOM (EP/J00796X/1).

\appendix

\section{Application of the formulas to specific cases\label{Appendix}}

In this appendix we specify our formulas to the cases of standard Bell
detection (balanced beam-splitter) and heterodyne detection (balanced
beam-splitter with a vacuum at one of the input ports). First we consider the
ideal case of unit quantum efficiencies for the detectors, i.e., we specify
the formula of Eq.~(\ref{output_FORMULA}). Then, we extend the results to the
case of arbitrary quantum efficiencies, which corresponds to applying the
formula of Eq.~(\ref{newInOut}).

\subsection{Standard Bell detection}

We achieve the formula for standard Bell detection by setting $T=1/2$
(balanced beam splitter). In this case, we have%
\begin{equation}
\gamma_{1}=\frac{1}{2}(\beta_{1}+\beta_{1}^{\prime})-\delta_{1},~\gamma
_{2}=\frac{1}{2}(\beta_{2}+\beta_{2}^{\prime})+\delta_{2},
\end{equation}
and $\gamma_{3}=\frac{1}{2}(\beta_{3}^{\prime}-\beta_{3}-\delta_{3}+\delta
_{4})$. Compactly, the $\boldsymbol{\gamma}$-matrix takes the form%
\begin{equation}
\boldsymbol{\gamma}=\frac{1}{2}\left(  \mathbf{ZB}_{1}\mathbf{Z}%
+\mathbf{B}_{2}-\mathbf{ZD}-\mathbf{D}^{T}\mathbf{Z}\right)  ~,\label{stGAMMA}%
\end{equation}
where $\mathbf{Z}$ is given in Eq.~(\ref{ReflectionMAT}). The $\mathbf{K}%
$-matrices can be simplified too. In fact, we get%
\begin{align}
\mathbf{K}_{11} &  =\frac{1}{2}\left(
\begin{array}
[c]{cc}%
\gamma_{2} & \gamma_{3}\\
\gamma_{3} & \gamma_{1}%
\end{array}
\right)  =\frac{1}{2}\mathbf{X}_{1}^{T}\boldsymbol{\gamma}\mathbf{X}%
_{1},\label{K11}\\
\mathbf{K}_{22} &  =\frac{1}{2}\left(
\begin{array}
[c]{cc}%
\gamma_{2} & -\gamma_{3}\\
-\gamma_{3} & \gamma_{1}%
\end{array}
\right)  =\frac{1}{2}\mathbf{X}_{2}^{T}\boldsymbol{\gamma}\mathbf{X}_{2}\\
\mathbf{K}_{12} &  =\mathbf{K}_{21}^{T}=\frac{1}{2}\left(
\begin{array}
[c]{cc}%
-\gamma_{2} & \gamma_{3}\\
-\gamma_{3} & \gamma_{1}%
\end{array}
\right)  =\frac{1}{2}\mathbf{X}_{1}^{T}\boldsymbol{\gamma}\mathbf{X}_{2}~,
\end{align}
where%
\begin{equation}
\mathbf{X}_{1}:=\left(
\begin{array}
[c]{cc}
& 1\\
1 &
\end{array}
\right)  ,~\mathbf{X}_{2}:=\left(
\begin{array}
[c]{cc}
& 1\\
-1 &
\end{array}
\right)  =\boldsymbol{\Omega}~.\label{XeOM}%
\end{equation}
Since $\mathbf{K}_{ij}=\mathbf{X}_{i}^{T}\boldsymbol{\gamma}\mathbf{X}_{j}/2$,
the formula of Eq.~(\ref{output_FORMULA}) becomes%
\begin{equation}
\mathbf{V}_{out}=\mathbf{A}-\frac{1}{2\det\boldsymbol{\gamma}}\sum_{i,j=1}%
^{2}\mathbf{C}_{i}(\mathbf{X}_{i}^{T}\boldsymbol{\gamma}\mathbf{X}%
_{j})\mathbf{C}_{j}^{T}~,\label{stBELL}%
\end{equation}
where the $\boldsymbol{\gamma}$-matrix is given by Eq.~(\ref{stGAMMA}).

It is straightforward to generalize the formula of Eq.~(\ref{stBELL}) to the
case of arbitrary quantum efficiencies $\eta$ and $\eta^{\prime}$ for the
homodyne detectors. In fact, it is sufficient to replace
\begin{equation}
\boldsymbol{\gamma}\rightarrow\boldsymbol{\gamma}(\eta,\eta^{\prime
})=\boldsymbol{\gamma}+\boldsymbol{\Phi}(\eta,\eta^{\prime})~,\label{replaAPP}%
\end{equation}
with $\boldsymbol{\gamma}$ given in Eq.~(\ref{stGAMMA}) and $\boldsymbol{\Phi
}(\eta,\eta^{\prime})$ given in Eq.~(\ref{FiGrande}).

\subsection{Heterodyne detection}

We achieve the heterodyne detection of the $(n+1)^{th}$\ mode by setting
$T=1/2$ (balanced beam splitter) and considering the last mode in a vacuum
state. As a matter of fact, this is equivalent to a standard Bell detection
where the last mode is in a vacuum. Thus, we have%
\begin{equation}
\mathbf{B}^{(2)}=\left(
\begin{array}
[c]{cc}%
\mathbf{B}_{1} & \mathbf{0}\\
\mathbf{0} & \mathbf{I}%
\end{array}
\right)  ~,~\mathbf{C}_{2}=\mathbf{0~,}%
\end{equation}
so that the global input CM is equal to%
\begin{equation}
\mathbf{V}_{in}=\left(
\begin{array}
[c]{cc}%
\mathbf{A} & \mathbf{C}_{1}\\
\mathbf{C}_{1}^{T} & \mathbf{B}_{1}%
\end{array}
\right)  \oplus\mathbf{I~.}%
\end{equation}
By setting $\mathbf{B}_{2}=\mathbf{I}$ and $\mathbf{D=0}$ in
Eq.~(\ref{stGAMMA}), we derive the expression of the $\boldsymbol{\gamma}%
$-matrix, which is given by%
\begin{equation}
\boldsymbol{\gamma}=\frac{1}{2}\left(  \mathbf{ZB}_{1}\mathbf{Z+I}\right)
=\frac{1}{2}\left(
\begin{array}
[c]{cc}%
\beta_{1}+1 & -\beta_{3}\\
-\beta_{3} & \beta_{2}+1
\end{array}
\right)  ~.\label{gammaHET}%
\end{equation}
This matrix has determinant%
\begin{equation}
\det\boldsymbol{\gamma=}\frac{\theta_{1}}{4},~\theta_{1}:=\det\mathbf{B}%
_{1}+\mathrm{Tr}\mathbf{B}_{1}+1~.\label{detGhet}%
\end{equation}
Since $\mathbf{C}_{2}=\mathbf{0}$, the sum in Eq.~(\ref{stBELL}) contains only
the term with $i=j=1$, i.e.,%
\begin{equation}
\mathbf{V}_{out}=\mathbf{A}-\frac{1}{2\det\boldsymbol{\gamma}}\mathbf{C}%
_{1}(\mathbf{X}_{1}^{T}\boldsymbol{\gamma}\mathbf{X}_{1})\mathbf{C}_{1}%
^{T}~.\label{vfinH}%
\end{equation}
Now, we can easily check that%
\begin{align}
\mathbf{X}_{1}^{T}\boldsymbol{\gamma}\mathbf{X}_{1} &  =\frac{1}{2}%
\mathbf{X}_{1}^{T}\left(  \mathbf{ZB}_{1}\mathbf{Z+I}\right)  \mathbf{X}%
_{1}\nonumber\\
&  =\frac{1}{2}\left(  \boldsymbol{\Omega}^{T}\mathbf{B}_{1}\boldsymbol{\Omega
}\mathbf{+I}\right)  =\frac{1}{2}\left(  \boldsymbol{\Omega}\mathbf{B}%
_{1}\boldsymbol{\Omega}^{T}\mathbf{+I}\right)  ~.\label{xGx}%
\end{align}
By using Eqs.~(\ref{detGhet}) and~(\ref{xGx}) into Eq.~(\ref{vfinH}), we get%
\begin{equation}
\mathbf{V}_{out}=\mathbf{A}-\frac{1}{\theta_{1}}\mathbf{C}_{1}\left(
\boldsymbol{\Omega}\mathbf{B}_{1}\boldsymbol{\Omega}^{T}\mathbf{+I}\right)
\mathbf{C}_{1}^{T}~.\label{HETformu1}%
\end{equation}
Similarly, if we heterodyne the $(n+2)^{th}$ mode with the $(n+1)^{th}$ mode
being the ancillary vacuum mode, we get%
\begin{equation}
\mathbf{V}_{out}=\mathbf{A}-\frac{1}{\theta_{2}}\mathbf{C}_{2}%
(\boldsymbol{\Omega}\mathbf{B}_{2}\boldsymbol{\Omega}^{T}+\mathbf{I}%
)\mathbf{C}_{2}^{T}~,
\end{equation}
with
\begin{equation}
\theta_{2}:=\det\mathbf{B}_{2}+\mathrm{Tr}\mathbf{B}_{2}+1~.
\end{equation}
This formula for the heterodyne detection coincides with that given in
Ref.~\cite{RMP} (without an explicit proof).

In this case too, we can easily extend the results to non-unit quantum
efficiencies, $\eta$ and $\eta^{\prime}$, for the homodyne detectors. We need
to perform the replacement $\boldsymbol{\gamma}\rightarrow\boldsymbol{\gamma
}+\boldsymbol{\Phi}(\eta,\eta^{\prime})$ in Eq.~(\ref{vfinH}). First note
that
\begin{equation}
\mathbf{X}_{1}^{T}(\boldsymbol{\gamma}+\boldsymbol{\Phi})\mathbf{X}_{1}%
=\frac{1}{2}[\boldsymbol{\Omega}(\mathbf{B}_{1}+2\boldsymbol{\Phi
})\boldsymbol{\Omega}^{T}\mathbf{+I}]~.\label{eq1FI}%
\end{equation}
Then, for the determinant we can write
\begin{equation}
\det(\boldsymbol{\gamma}+\boldsymbol{\Phi})=\det\boldsymbol{\gamma}%
+\det\boldsymbol{\Phi}+\mathrm{Tr}\left(  \boldsymbol{\Omega\Phi\Omega}%
^{T}\boldsymbol{\gamma}\right)  ~,\label{detSVIL}%
\end{equation}
which is an equality valid in general for any symmetric matrix
$\boldsymbol{\gamma}$ and diagonal matrix $\boldsymbol{\Phi}$. Now using
Eq.~(\ref{detGhet}) in Eq.~(\ref{detSVIL}), we get%
\begin{align}
\det(\boldsymbol{\gamma}+\boldsymbol{\Phi})  & =\frac{\theta_{1}%
+4\det\boldsymbol{\Phi}+2\mathrm{Tr}\boldsymbol{\Phi}+2\mathrm{Tr}\left(
\boldsymbol{\Omega\Phi\Omega}^{T}\mathbf{B}_{1}\right)  }{4}\nonumber\\
& :=\frac{\theta_{1}(\eta,\eta^{\prime})}{4}~.\label{theta1app}%
\end{align}
Finally, using Eqs.~(\ref{eq1FI}) and~(\ref{theta1app}), we get%
\begin{equation}
\mathbf{V}_{out}(\eta,\eta^{\prime})=\mathbf{A}-\frac{1}{\theta_{1}(\eta
,\eta^{\prime})}\mathbf{C}_{1}[\boldsymbol{\Omega}(\mathbf{B}_{1}%
+2\boldsymbol{\Phi})\boldsymbol{\Omega}^{T}\mathbf{+I}]\mathbf{C}_{1}^{T}~,
\end{equation}
which is the generalization of Eq.~(\ref{HETformu1}) to arbitrary quantum efficiencies.


\begin{thebibliography}{99}
\bibitem {Proof}Proof is trivial. Set $\mathbf{M}=x\boldsymbol{\Pi}$ and
$\mathbf{G}=\left(
\begin{array}
[c]{cc}%
a & c\\
d & b
\end{array}
\right)  $. Then, $\mathbf{H}=\left(
\begin{array}
[c]{cc}%
ax-1 & cx\\
0 & -1
\end{array}
\right)  $ and $r=(ax-1)^{2}+c^{2}x^{2}+1$. The gradient of $r$ is given by
$\frac{\partial r}{\partial c}=2x^{2}c$ and $\frac{\partial r}{\partial
a}=2x(ax-1)$ which is zero for $c=0$ and $a=x^{-1}$ (since $x\neq0$). Finally,
we can freely choose $b=d=0$, so that $\mathbf{G}=x^{-1}\boldsymbol{\Pi}$.
\end{thebibliography}
\end{document}